\documentclass[a4paper, onecolumn, preprint]{revtex4}
\usepackage{amsmath}
\usepackage{amsfonts}
\usepackage{graphicx}
\usepackage{color}
\usepackage{lscape}
\usepackage{amssymb}

\begin{document}

\title{Establishing a microscopic model for nonfullerene organic solar cells: Self-accumulation effect of charges}
\author{Yao Yao}
\email{yaoyao2016@scut.edu.cn}
\affiliation{Department of Physics and State Key Laboratory of Luminescent Materials and Devices, South China University of Technology, Guangzhou 510640, China}

\begin{abstract}
A one-dimensional many-body model is established to mimic the charge distribution and dynamics in nonfullerene organic solar cells. Two essential issues are taken into account in the model: The alternating donor and acceptor structure and the local imbalance of the intrinsic electrons and holes. The alternating structure is beneficial for the direct generation of charge transfer state which enhances the local imbalance of intrinsic charges. The most remarkable outcome of the model is that, due to the strong Coulomb attractive potential energy, the intrinsic charges in the cells are self-accumulated in a small spatial region. Outside the self-accumulation region, the charge density vanishes so that the recombination is regarded to be largely suppressed. The photogenerated electrons are subsequently observed to spread freely outside the self-accumulation region implying the Coulomb attraction does not matter in the ultrafast charge separation dynamics. These findings enable an appealing understanding of the high performance of emerging nonfullerene cells, and the designing rules of molecules and devices are then comprehensively discussed.
\end{abstract}

\clearpage
\maketitle

\section{Introduction}

Massive researches were devoted to nonfullerene organic solar cells (OSCs) in recent years due to their rapidly developing performance \cite{ReviewH,Chen1,NFA1,NFA2,Zhan2,Zhan1,Zhan3,Zhan4,HouAM,NFA3,HouJACS,NFA4,LiCM,HouCJC,P14,Zhan5,Ter0}. The up-to-date power conversion efficiency has achieved 17.3$\%$ \cite{Chen1}, exceeding the long-term bottleneck of fullerene-based cells. The substantial progress benefits from two novel strategies. Firstly, several nonfullerene acceptors (NFAs) are utilized, such as PDI, ITIC, N2200 and their derivatives \cite{Ter0}. These NFAs normally possess much smaller band gap than that of fullerene so that the optical absorption of nonfullerene cells are much better than that of fullerene-based cells. Secondly, tandem structures are extensively employed so that the absorption of cells effectively covers the entire solar spectrum which further improves the efficiency \cite{Chen1,NFA4,P14}. Despite the high device performance, the tandem structure is too complicated for a practical fabrication limiting its potential commercial applications. Researchers thus put great effort into the study of (cascade) multi-component blends, such as the donor/acceptor (D/A) binary blends and also the ternary ones with more complicated combinations of D and A components \cite{Ter0,Ter1,Ter2,Ter3,Ter4,Ter5,Ter6,Ter7,Ter8,Ter9,Ter10,Ter11,Ter12,Ter13,Ter14}. The multi-component blends, both in all-polymer \cite{Ter1} and all-small-molecule \cite{LiCM} cells, perform as perfect as the tandem cells exhibiting promising application potentials in a very near future.

The physical origin of driving force for the charge separation turned out to be a long-standing puzzle in OSCs \cite{Review}. The relatively low dielectric constant of organic materials results in a strong Coulomb attractive potential energy between electron and hole which is much stronger than that of the thermal fluctuation and the built-in electric field, leading to the difficult dissociation of photogenerated excitons. Therefore, in a long period of investigations on the OSCs, the acceptor is consistently chosen to be fullerene and its derivatives. The special molecular structure of fullerene gives rise to sufficient available orbits for electrons and thus the strong electron affinity. In other words, in order to achieve enough driving force for the efficient charge separation, the fullerene-based cells compromise the optical absorption of acceptor by using fullerene with wide energy gap. Due to the fixed gap of fullerene acceptors, the choice of donors are greatly limited. In order to improve the absorption and thus the short-circuit current ($J_{\rm sc}$), the donors should be of low energy gap, and unfortunately the energy offset between the donor and acceptor is subsequently reduced lowering the driving force of charge separation and thus the open-circuit voltage ($V_{\rm oc}$). These two competing factors, $J_{\rm sc}$ and $V_{\rm oc}$, constitute the bottleneck of the cell efficiency. Furthermore, it is stated that the wavefunction of electron tends to be delocalized among the fullerene molecules and the charge separated states have got larger density than that of local excitons \cite{Delo1,Delo2,Delo3,Delo4,Entropy,Yao16}. These two arguments are frequently mentioned to explain the mechanism of charge generation in fullerene-based cells.
From these points of view, materials with better crystallinity or purity should perform better than amorphous ones. In the real situations, however, perylene derivatives with good  crystallinity normally exhibit worse performance than other NFAs and ternary structures act as well as binary ones \cite{Ter0}, suggesting the delocalization mechanism does partly not work in these systems.

Nonfullerene cells extensively employ the ``seem-to-be-poor" tandem or ternary structures and perform very well \cite{Ter0}. They seem to be poor because the NFAs always have relatively low band gap so that in these structures the energy offset between donor and acceptor and thus the driving force should be even smaller than that in fullerene-based cells. A recent experiment has demonstrated that the driving force in nonfullerene cells is much smaller than that in fullerene-based cells \cite{NE}, which implies the charge separation itself is a spontaneous process and does not need an energetic driving force. In the same experimental research, the $J_{\rm sc}$ is found to be largely improved by finely tuning the molecular structures while the $V_{\rm oc}$ is concurrently increased manifesting contradictory to the normal sense. As a consequence, the great achievements of nonfullerene cells are currently challenging all the conventional theories for the charge separation in OSCs, and a comprehensive theoretical model applicable in this new kind of cells is in high demand.

In order to rationally build a theoretical model that appropriately mimicking the major physics in nonfullerene cells, let us first briefly summarize the essential experimental findings up to date. First of all, due to the efficient absorption of photons in both donor and acceptor, the film of device is not necessary to be very thick, so that the mobilities and thus the single-carrier transport no longer play the essential role. The charge separation still serves as the key issue for the device performance. Secondly, different from the fullerene molecule, the NFAs always comprise complicated functional units which must be carefully considered. For example, the end groups manifest strong charge affinity, and the side groups on the other hand shape the whole blend to be an alternating structure of D and A units. Lastly, the polaron pair state as a component of the photogenerated excited state has been demonstrated to exist \cite{PP1,PP2}. The model in the present work will be established on the basis of these issues. The paper is organized as follows. In Section II we discuss the main features of nonfullerene cells and propose a model Hamiltonian to mimicking these cells. In Section III the calculating results are present, and the self-accumulation effect and the ultrafast charge separation are investigated. The conclusion and outlook are drawn in Section IV.

\section{Microscopic model}

\begin{figure}
\includegraphics[scale=0.4]{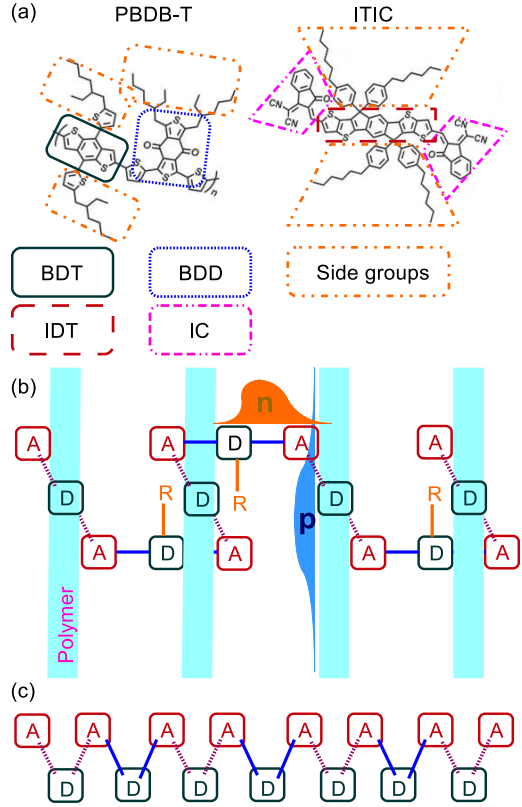}
\caption{(a) Chemical structures of PBDB-T and ITIC. (b) Model of an alternating structure of D and A units. The A--D--A structure of acceptor molecules is sandwiched in two polymer chains. Negative charges denoted by `n' are assumed to reside on the A unit of ITIC while positive charges denoted by `p' distribute along the polymer chain. `R' represents the side group. The solid lines represent the intramolecular bonds and the dashed lines denote the intermolecular interactions ($\pi-\pi$ stacking). (c) Schematic of the many-body tight-binding model abstracted from the real materials.}\label{Fig1}
\end{figure}

In this work, we take the blend of PBDB-T and ITIC as instance to build our microscopic model \cite{Chem}. PBDB-T is a conjugated polymer donor composing of a D--A structure with D unit being BDT and A unit being BDD \cite{HouAM}. ITIC was firstly synthesized in 2015 and quickly became a benchmarking small-molecule NFA \cite{Zhan2}. It has an A--D--A structure with a core D unit being IDT and two end units of A being IC. The chemical structures of the molecules are displayed in Fig.~\ref{Fig1}(a). In this blend, the major electron-withdrawing groups should be the end groups of acceptor molecules, i.e. cyano- and/or ketone-group in ICs.

As the component of nonfullerene OSCs, molecules usually comprise three functional units \cite{Zhan5}. (1) The core groups, such as the thiophene, the perylene, the fused ring in ITIC and the backbone in polymers, dedicate the main conjugated orbits for electron and/or hole polarons. (2) The end groups with different electron affinity (and possibly some paired radicals \cite{Huang,Radical}) induce local charge dipoles which is the intrinsic charge transfer effect, and the imbalanced spatial distribution of negative and positive charges break the local neutrality in the device. For example, the fluorine or the cyanogroup has strong ability of electron attraction, and experiments have found that substitution of these groups can largely enhance the device performance \cite{HouJACS,LiCM}. (3) The side groups with poor conductivity shape the special structure of the blend. Interestingly, one can find in Fig.~\ref{Fig1}(a), there are long side groups on the D units in both PBDB-T and ITIC. These side groups act as struts to brace the whole molecules, such that the D unit in PBDB-T does not directly contact to the D unit in ITIC \cite{ReviewH,private}. This means the blend should be of a repeating structure of D$\cdots$A--D--A$\cdots$, with `$\cdots$' indicating the intermolecular contact and `--' being the intramolecular bond. It is worth noting that, the side group engineering is commonly presented in nonfullerene cells, implying the alternating structure of D and A units emerges as an essential feature for the high performance \cite{Zhan5}.

As the first important issue inputted in our model, the structure of PBDB-T/ITIC heterojunction could be abstractly sketched as follows. The polymer chains of PBDB-T are parallel to each other, and the A--D--A structure of ITIC is sandwiched in two polymers with $\pi-\pi$ stacking between the D unit of PBDB-T and A unit of ITIC. The realistic situation could be slightly different because the polymers are always disordered and entangled, but this does not matter because the parallelism is not the central point of our simplified model. The side groups labeled by `R' in the D unit of ITIC separate the D units and shape a quasi-one-dimensional (1D) structure.

The dimensionality is an extensively discussed issue in OSCs. In the traditional fullerene-based cells, people rationally thought that the fullerene molecules have a three-dimensional structure such that the entropy for free electron states in fullerene is much larger than that in polymer donors which have got low dimensionality \cite{Review}. Significantly different from fullerene, most commonly-used small-molecule NFAs possess the planar structure and spatial anisotropy, suggesting their dimensionality is smaller than 3, or even than 2. There are two evidences. One is that the molecular orientation can largely influence the device performance of nonfullerene cells as reviewed by Hou et al. \cite{ReviewH}. The other is the weak electron-spin resonance signal in ITIC showing that the two radicals in ITIC form a singlet pair and excluding the high-dimensional structure in which the electron is more free \cite{Radical}. By these considerations, we believe our model is applicable to most planar NFAs. In addition, although the bulk heterojunction structure of OSCs leads to amorphous phase, there is a 1D percolating pathway for the charge transport. We thus borrow this picture to mimic the nonfullerene solar cells with a quasi-1D microscopic model.

The second essential point introduced in the model is the imbalanced charges in the system. There are two origins of charges in the cells. The first is the {\it intrinsic charge} in the system self-doped by either electropolar (end) groups or radicals which can be self-accumulated in a small region as addressed below \cite{Huang}. The second is the {\it photogenerated charge}. The photogenerated excited states have been demonstrated to comprise two components \cite{PP1,PP2}. One is the usual Frenkel exciton with local charge neutrality, and more importantly there is another elementary excitation, the charge transfer state (or the polaron pair state) \cite{PPnote}. The charge transfer state consists of two spatially separated polarons with opposite charges. The excitations in the nonfullerene cells are thus the combinations of Frenkel excitons and polarons. We realize that the different electron affinity and the alternating D and A structure can intuitively enhance the generation rate of polarons.

For convenience of expression, in the following we uniformly call the (intrinsic/photogenerated) negative charges as (intrinsic/photogenerated) electrons and positive charges as holes. To avoid confusing, readers should bear in mind that intrinsic and photogenerated electrons are indistinguishable particles in the theoretical modelings, but they stem from different origination. We assume in our situation the intrinsic electrons locally reside on the A unit of ITIC while the intrinsic holes distribute along the polymer chain due to the conjugation of polymers, as drawn in Fig.~\ref{Fig1}(b). It means some intrinsic holes distribute out of the range of the 1D alternating structure that we are investigating. As a result, the local density of electrons on the 1D alternating structure should be larger than that of holes. It is apparent that the inverse situation could produce qualitatively same conclusions. Furthermore, as we are mainly consider the dynamics on an ultrafast timescale, the recombination terms will not be involved in the model. We will argue that due to the self-accumulation effect, the recombination in the realistic cells could be largely suppressed.

We are now on the stage to build the 1D tight-binding model with the topological structure schematized in Fig.~\ref{Fig1}(b). The solid lines in Fig.~\ref{Fig1}(b) represent the intramolecular covalence bonds in the ITIC molecules and the dashed lines denote the intermolecular interactions between PBDB-T and ITIC. Via the bridge effect of opposite units \cite{bridge1,bridge2}, holes move among the D units while electrons move among the A units, respectively. As a normal consideration in tight-binding models, we assume the hopping of hole or electron can only occur between the nearest D or A unit, respectively, and for simplicity we neglect the disorder of the hopping constant in the present work, which will be held as a future subject. Electron and hole residing on nearest units will feel a strong attractive potential energy which could be different for intra- and inter-molecular bond. We do not consider the long-range attraction because of the electrostatic screening of the opposite charges. It is worth noting that, if the long-range attraction was taken into account, the main conclusion in this work, i.e. the self-accumulation effect, would be further enhanced.

The Hamiltonian of the 1D alternating structure as schematized in Fig.~\ref{Fig1}(c) is thus written as
\begin{eqnarray}
H&=&-\sum_{j}(t_{\rm e}c^\dagger_{2j}c_{2j+2}+t_{\rm h}d^\dagger_{2j+1}d_{2j+3}+{\rm h.c.})\nonumber\\&+&\sum_{j}U_jc^\dagger_{2j}c_{2j}d^\dagger_{2j+1}d_{2j+1},
\end{eqnarray}
where $c^\dagger_{2j} (c_{2j})$ creates (annihilates) an electron on $2j$-th site (unit); $d^\dagger_{2j+1} (d_{2j+1})$ creates (annihilates) a hole on $(2j+1)$-th site; $t_{\rm e/h}$ is the hopping constant for electron/hole and in this work we set them to 100meV (on the order of vibrational frequencies); $U_j$ is the attractive potential energy between nearest electron and hole. In order to distinguish the intra- and inter-molecular interactions, $U_j$ for odd and even $j$ will be set to different values $U_1$ and $U_2$, respectively. For simplicity, we set $U_1$ to be $-400$meV and adjust $U_2$ in the practical computations. It is reasonable that $U_1$ is four times larger than the hopping constant, since the dielectric constant in organic molecules is as small as 3 and the Coulomb interaction is relatively strong. We note here that, in realistic materials the results could be quantitatively different with those presented in this work because of the distinct parameters, but the qualitative conclusions are still valid.

One would be noting that, if we consider the electron as a spin up and the hole as a spin down, the model is nothing but a 1D extended Hubbard model with negative Hubbard $U$ and next-nearest interactions. The features in the Luttinger liquid theory could be safely applied to the model. For example, with negative $U$ there is a phase named ``phase separation" in which the charges are accumulated in a local region \cite{Lin}. The self-accumulation effect shown below actually stems from this effect. In addition, the Hubbard model and the Luttinger liquid theory have been applied to the benzene and fullerene systems \cite{Lut1,Lut2}, but those are solely related to single-component systems. Our model is more generic for multi-component cells.

\section{Results}

In this section, we show the results calculated by the static and adaptive time-dependent density matrix renormalization group algorithm \cite{SCS,Yao08}. The total number of sites in the system is set to 128 and the truncation number of states is 64. The key parameters are the total number of electron and hole in the system, namely $N_{\rm e}$ and $N_{\rm h}$, respectively. Without loss of generality, we mainly study the case that the electron number is larger than that of hole in this work.

\subsection{Self-accumulation effect of intrinsic charges}

\begin{figure}
\includegraphics[scale=0.65]{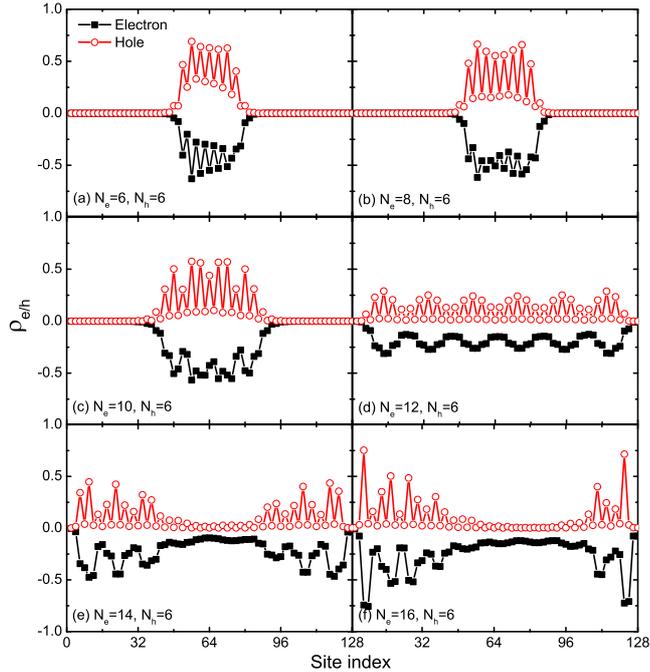}
\caption{Charge densities $\rho_{\rm e/h}$ on the lattice for $N_{\rm h}=6$, $U_2=-200$meV and six sets of $N_{\rm e}$.}\label{Fig2}
\end{figure}

\begin{figure}
\includegraphics[scale=0.3]{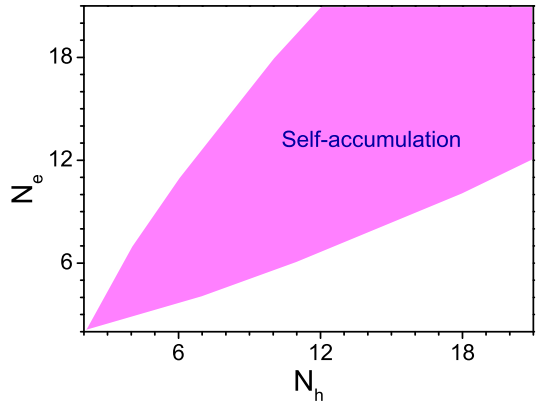}
\caption{Phase diagram of the model for $U_2=-200$meV and different sets of $N_{\rm e}$ and $N_{\rm h}$. The purple area illustrates the self-accumulation phase.}\label{Fig3}
\end{figure}

We first study the properties of the intrinsic charges of the system without photoexcitation. The local charge densities $\rho_{\rm e/h}(x,t)$ for electron (minus sign) or hole (positive sign) with $x$ being the site index are calculated with $N_{\rm h}=6$, $U_2=-200$meV and different $N_{\rm e}$, as shown in Fig.~\ref{Fig2}. It is found that, when $N_{\rm e}\leq$10 the intrinsic charges are accumulated in a small spatial region of the lattice, and outside the region the charge densities are completely vanishing. For example, in the case $N_{\rm e}=8$ the charge densities are accumulated within 40 sites. Holes are spaced out over 20 sites while electrons are approximately evenly distributed, since now the total number of electron is larger than that of hole. As stated, this is the feature of the phase separation of intrinsic charges \cite{Lin}. The strong Coulomb attraction between electron and hole results in the accumulation. In organic materials, due to the small dielectric constant the Coulomb attraction is always sufficiently strong so that the accumulation found here could be spontaneously present in realistic cases. Intrinsic charges in the vicinity are easily absorbed into the accumulation region making the vicinity region empty. We call this effect to be ``self-accumulation", which is one of the essential consequences of the present work. When $N_{\rm e}>$10 the self-accumulation effect breaks down and the system enters into the phase of charge density wave.

We can parallel calculate the cases of other values of $N_{\rm h}$, and Fig.~\ref{Fig3} displays a phase diagram to figure out the parameter regimes for the appearance of self-accumulation effect. Obviously, when the intrinsic electrons and holes are nearly equivalent, namely $N_{\rm e}\simeq N_{\rm h}$, the self-accumulation effect emerges as expected. The phase boundaries are close to linear and located at around $N_{\rm e}/N_{\rm h}=N_{\rm h}/N_{\rm e}\simeq 1.75$. One can see that, the parameter regime for the self-accumulation effect is quite flexible. It is not necessary that the electron density is extremely larger than that of hole, so a number of organic materials can be chosen as candidates for the nonfullerene cells. As discussed later, however, if $N_{\rm e}$ is too close to $N_{\rm h}$ the photogenerated charges are not easy to be dissociated. Subsequently, we conclude here an optimal condition for an efficient nonfullerene cell is that $N_{\rm e}/N_{\rm h}$ or $N_{\rm h}/N_{\rm e}$ lies between 1.5 and 2.0, depending on the specific material parameters.

As a critical consequence, the self-accumulation effect of intrinsic charges stems from the alternating D and A structure. As aforementioned, this effect leads to the suppression of charge recombination, since there are few charges outside the self-accumulation region. It is well accepted that the nonfullerene cells manifests good performance, but people do not understand why the ultrahigh recombination loss of $V_{\rm oc}$ observed in traditional fullerene-based cells is suppressed in this new kind of cells \cite{loss1,loss2}. Herein, we provide a possible explanation of this issue by uncovering the self-accumulation effect of intrinsic charges.

\subsection{Spread of photogenerated charges outside the self-accumulation region}

\begin{figure}
\includegraphics[scale=1.1]{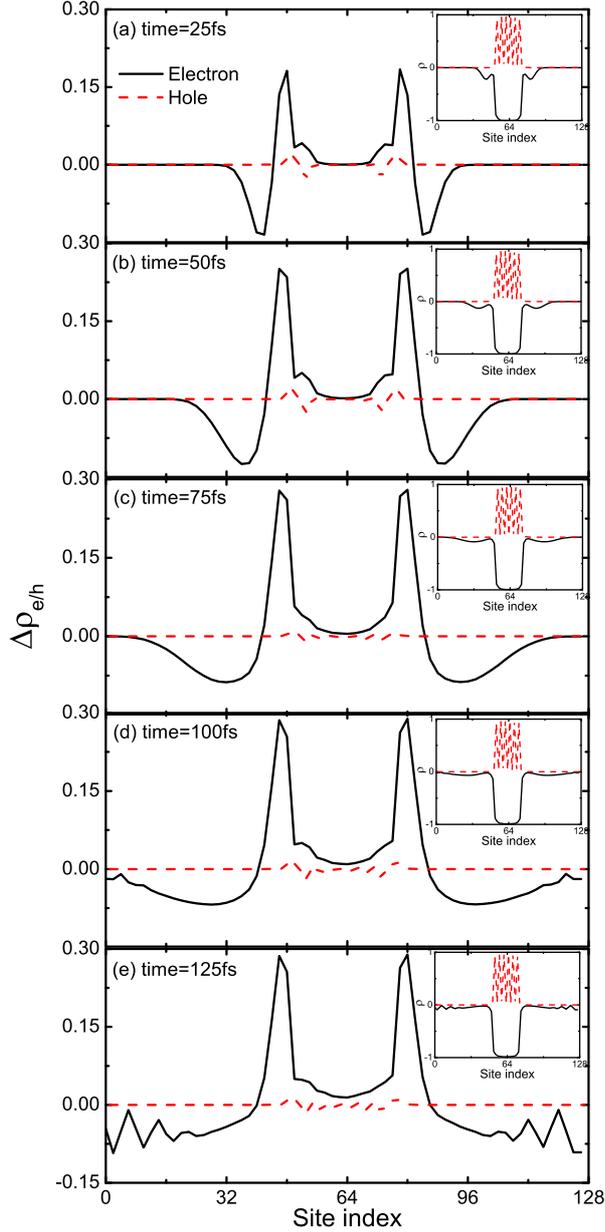}
\caption{Spatial distribution of $\Delta\rho_{\rm e/h}$ at five time points. Insets show the relevant distribution of charge densities $\rho_{\rm e/h}$. The parameters are $N_{\rm e}=14$, $N_{\rm h}=6$ and $U_2=-200$meV.}\label{Fig4}
\end{figure}

\begin{figure}
\includegraphics[scale=1.1]{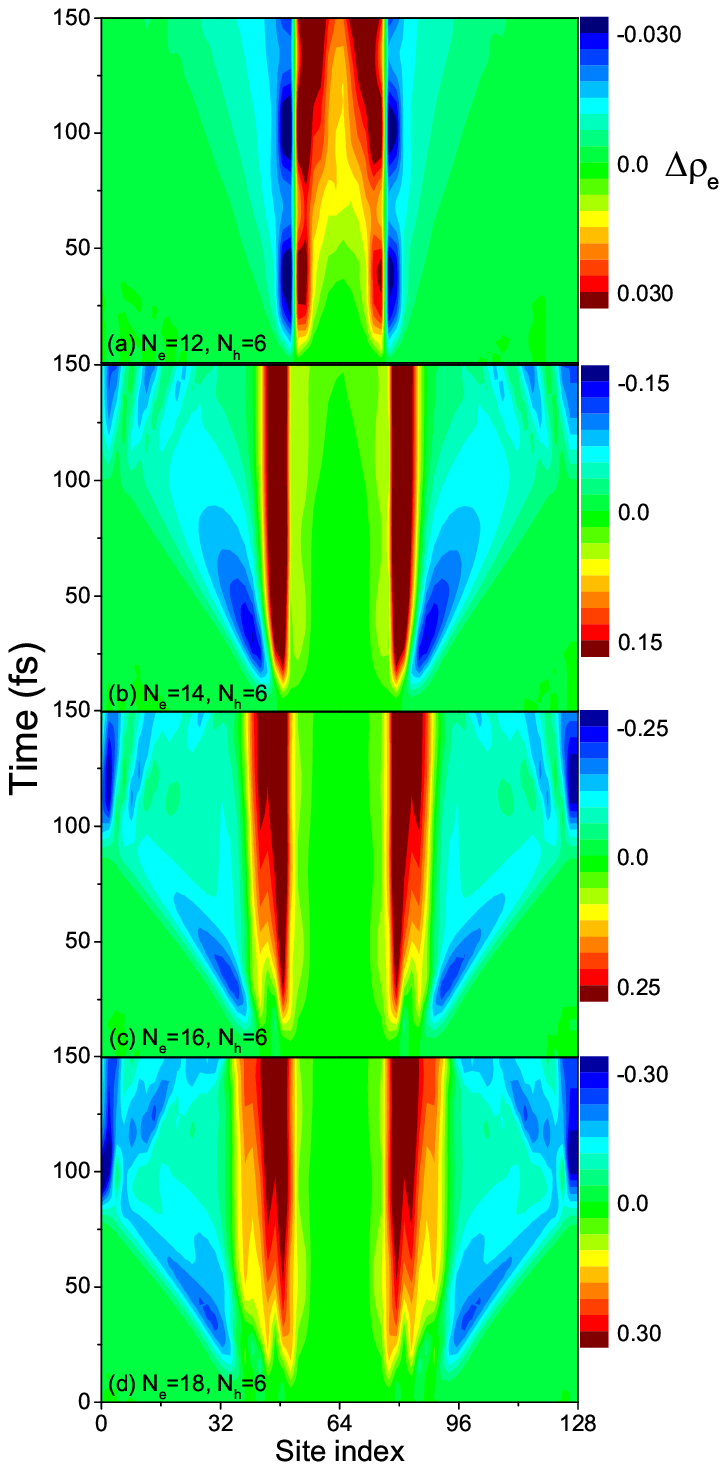}
\caption{Time evolution of $\Delta\rho_{\rm e}$ on the lattice for four sets of $N_{\rm e}$ and $N_{\rm h}=6$, $U_2=-200$meV.}\label{Fig5}
\end{figure}

\begin{figure}
\includegraphics[scale=1.1]{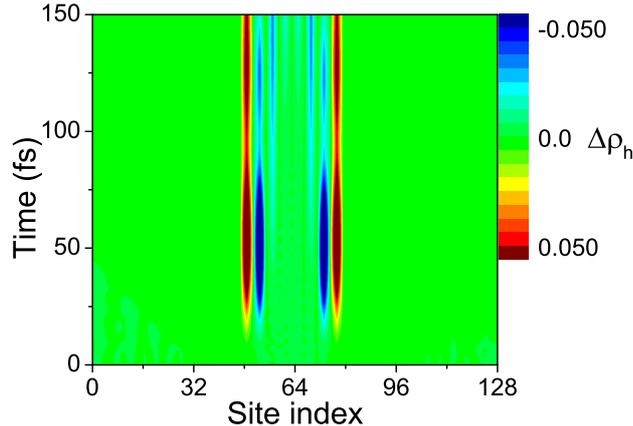}
\caption{Time evolution of $\Delta\rho_{\rm h}$ on the lattice for $N_{\rm e}=16$, $N_{\rm h}=6$ and $U_2=-200$meV.}\label{Fig6}
\end{figure}

\begin{figure}
\includegraphics[scale=1.1]{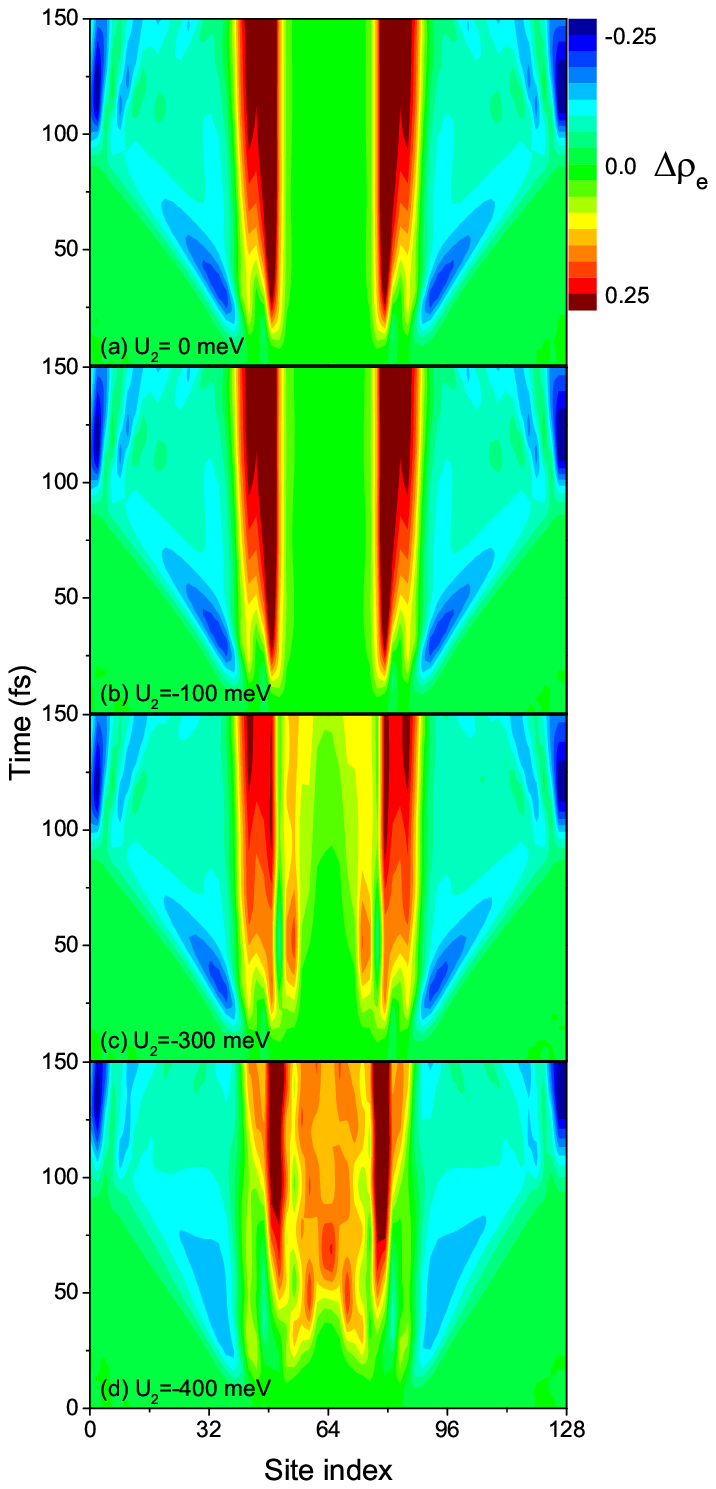}
\caption{Time evolution of $\Delta\rho_{\rm e}$ on the lattice for four sets of $U_2$ and $N_{\rm e}=16$, $N_{\rm h}=6$.}\label{Fig7}
\end{figure}

In this subsection, we calculate the dynamics upon photoexcitation. As shown in Fig.~\ref{Fig3}, for $N_{\rm h}=6$ the phase boundary of self accumulation locates at $N_{\rm e}\simeq 10$. For mimicking the physical situation of photoexcitation, we fix $N_{\rm h}$ to be 6 and set $N_{\rm e}$ to be larger than 10, which is the maximum number of intrinsic electrons for the presence of self accumulation as discussed in the last subsection, and the excess electrons over the 10 intrinsic electrons are thus regarded to be photogenerated. At time $\leq 0$ we add a balance potential energy to the system for both electron and hole with the form being $V(x)=-0.5\exp[-(x-x_{\rm c})^2/400]$ with $x_{\rm c}$ the center of the lattice. In this situation, the electrons and holes will be initially accumulated in the middle of the lattice. Once time $>0$ the potential is switched off to mimic the situation that the laser pulse is switched off and the photogenerated electrons starts to spread. We then calculate $\Delta\rho_{\rm e/h}[\equiv\rho_{\rm e/h}(t)-\rho_{\rm e/h}(0)]$ to quantify the spreading charges.

In Fig.~\ref{Fig4}, the spatial and temporal dependence of both $\Delta\rho_{\rm e/h}$ and $\rho_{\rm e/h}$ are shown with $N_{\rm e}=14$, $N_{\rm h}=6$ and $U_2=-200$meV. As expected, both the electron and hole are initially accumulated in the middle of the system due to the initial potential energy. After the initial potential is off, it is found that a part of electrons which can be regarded as photogenerated electrons quickly spread out to the two ends of the lattice. At around 100fs the photogenerated electron wavepackets travel over 40 sites and arrive in the ends, and then they will stay there as we do not consider the recombination loss. In a realistic situation, 40 sites are long enough for the electron-hole pair to be dissociated. In order to maintain the conservation of the particle number in the entire system, in the middle of the lattice two positive peaks are induced which do not move during the whole process. The summation of the density of spreading photogenerated electrons at 100fs equals to 1.89, which means about 2 electrons spread out of the self-accumulation region. On the other hand, one can find the hole density (red dashed lines in Fig.~\ref{Fig4}) does almost not change at all. This is surprising because we expected that the spreading electrons would be able to pull at least a few holes out due to the strong attraction among them. It suggests a new mechanism for the charge separation of photogenerated electron-hole pairs. As stated, when $N_{\rm e}\leq 10$ the intrinsic charges are self-accumulated, so that 10 electrons are recognized to be resident (intrinsic charges) self-doped by electropolar groups or radicals. The other electrons could therefore be regarded to be originated from photogeneration, and 2 of them spontaneously leave the self-accumulation region leading to an effective dissociation of the electron-hole pairs and ultrafast charge separation process. This effect turns out to be the second remarkable finding in this work.

In order to see the essential parameters affecting the motion of photogenerated electron and hole, we adjust the total number of electron $N_{\rm e}$ while keeping $N_{\rm h}$ to be 6. It is found in Fig.~\ref{Fig5}, when $N_{\rm e}=12$ the spread of photogenerated electron is extremely slow. When $N_{\rm e}\geq 14$, the spread becomes efficient, and the larger the $N_{\rm e}$ is, the faster the spread is. As a comparison, the time evolution of hole density is plotted in Fig.~\ref{Fig6}, where one can observe that the hole density does almost not spread in the entire process. Notice that in our model we do not consider any energy offsets and energetic driving forces, this could be regarded as a novel scenario of charge separation in the presence of strong Coulomb attraction between electron and hole. We also calculate the time evolution for different $U_2$ and the results are shown in Fig.~\ref{Fig7}. Following $U_2$ increasing, the region of self-accumulation becomes shrunk, while the motion of photogenerated electron outside the region is not affected.

As an additional remark, the ultrafast charge separation process presented here is completely different from that in the planar heterojunctions which is correlated with the long-range charge transfer state \cite{Yao16}. In the latter case, there is not an alternating structure so that the many-body model can not be applied and the self-accumulation effect can not be found. As we studied in the previous work, a nonlocal electron-phonon interaction is necessary to serve as the driving force for the charge separation \cite{Yao16}. In the present model, the electron-phonon interaction is not taken into account and the driving force turns out to be the many-body effect of electrons and holes.

\subsection{Photogenerated charge transport outside the self-accumulation region}

\begin{figure}
\includegraphics[scale=1.1]{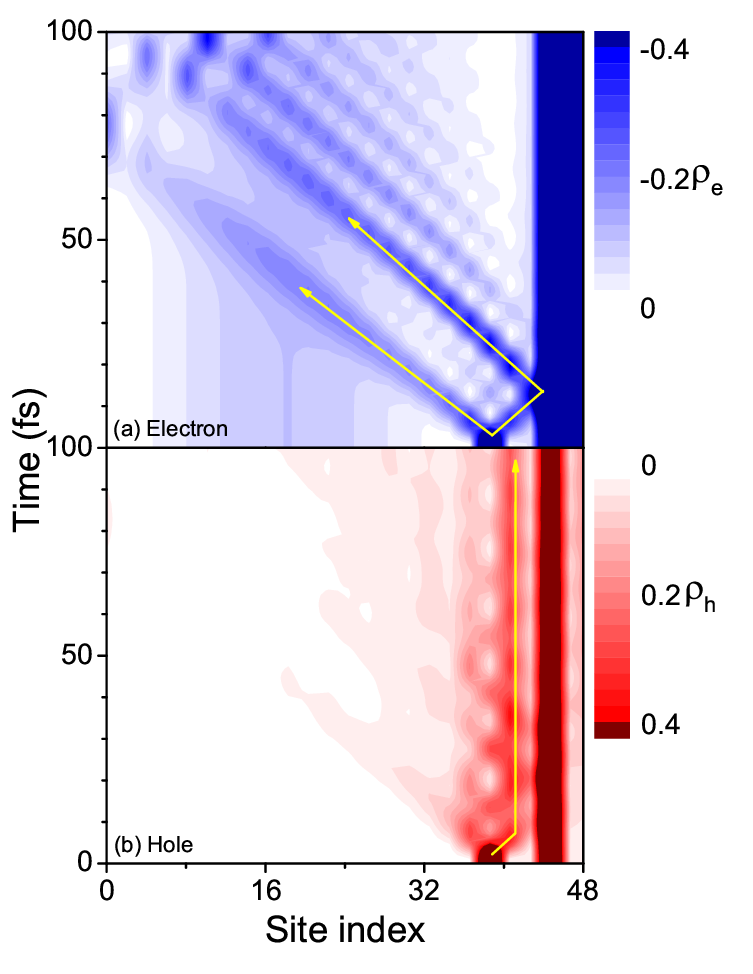}
\caption{Time evolution of $\rho_{\rm e/h}$ on the lattice for $N_{\rm e}=14$, $N_{\rm h}=6$, $U_2=-200$meV. Arrows indicate the paths of electrons and holes.}\label{Fig8}
\end{figure}

As the photogenerated electrons will spontaneously spread out of the self-accumulation region, one would be wondering what will happen when these photogenerated electrons encounter other self-accumulation region: Will they be absorbed by this region due to the strong Coulomb attraction? If the answer is yes, it means the recombination loss of photogenerated charges is still large in the cells. In order to mimic the physical scenario, in the presence of $V(x)$, an additional term $\tilde{V}_{\rm e/h}$ is initially acted on the site $x_c-25$ on which local electron/hole densities are induced. To distinguish the behavior of electron and hole, $\tilde{V}_{\rm e/h}$ is adjusted to be different values for electron and hole such that we can separately study the motion of photogenerated electron and hole and eliminate the influence of the interaction between them.

Fig.~\ref{Fig8} shows the evolution of $\rho_{\rm e/h}$ for $N_{\rm e}=14$, $N_{\rm h}=6$, $U_2=-200$meV. In order to be more focused, only 48 sites are displayed. As indicated by the arrows, there is a local excitation of electron/hole charge density outside the self-accumulation region, and following time evolving the charges split into two branches and spread to opposite directions. For electrons, the right-moving branch runs and crashes against the self-accumulation region, and quickly it rebounds to the left. It means the photogenerated electrons are not captured by the accumulated charges at all. This result suggests that the photogenerated electrons are nearly free and have small chance to be captured and recombined. On the other hand, the holes behave different as one can find in Fig.~\ref{Fig8}(b) that the majority hole densities are localized adhering to the self-accumulation region, implying that the holes are easy to be captured. This is easy to understand since we are studying the case that the electron is majority. As a result, so long as the local density of electron is larger than that of hole, they can be efficiently separated and the electron can be smoothly transported out of the active layer.

\section{Conclusion and outlook}

In summary, we have presented a 1D many-body microscopic model taking the strong Coulomb attraction between electron and hole into account. By analysis of the molecular structures of PBDB-T and ITIC, an alternating D and A structure is proposed which gives rise to efficient generation of charge transfer states. It is found that there is a self-accumulation region in the electron-rice case and outside the region there are few charges leading to free transport of photogenerated electrons. In a dynamical manner, the photogenerated electron is uncovered to be easily dissociated from the self-accumulation region. We realize that this effect could be applied to explain the underlying mechanism of ultrafast charge separation in nonfullerene OSCs.

As an outlook, we would like to discuss more on the designing rules of OSCs. As stated, fullerene molecules possess denser states for electrons and higher symmetry than other organic molecules, so that in a long history the fullerene dominated the battles among electron acceptors. Fullerene molecules however hold very obvious drawback, i.e. its energy gap is too wide to absorb the solar emission. This drawback constraints the choice of donor materials which must be of low energy gap and the choice of device structure in which the donor layer must be sufficient thick. As a result, both the $V_{\rm oc}$ and $J_{\rm sc}$ easily meet their bottlenecks in the fullerene-based cells. In the very initial studies of NFAs, the designing rule suitable for the fullerene is also partly applied to the NFAs slowering the developing progress.

Based upon the new insights in nonfullerene cells, the designing rules should also be modified to a large extent. As a conclusion, we here summarize two sufficient conditions for an efficient charge separation in the cells. (1) An alternating structure of D and A units is essential. The reason is twofold. Firstly, the D--A structure prefers the charge transfer states rather than Frenkel excitons, which means in this structure the charge transfer states are more easily to be generated than in other structures. Secondly, as found in this work, the D and A alternating structure gives rise to self-accumulation of charges implying outside the region there are few charges. As stated, this is critically important because the recombination outside the accumulated region is overwhelmingly weakened, such that the transporting electrons outside the accumulated region can be smoothly collected by the electrodes. This can be adopted to explain the reason that the $V_{\rm oc}$ loss is not significant in nonfullerene cells \cite{NE}. On the experimental side, the alternating structure could be achieved by either properly modulating the side groups or adding strong electropolar groups to firmly connect the D and A units. In our work, we take a 1D structure as instance, but a 2D structure still works because the 2D system also shares the features of the Luttinger liquid. (2) There has to be imbalanced local charge densities of intrinsic electron and hole, and in a local system with an equal number of electrons and holes the mechanism does not work at all. Apart from these two issues, the electronic energies (to form cascade structure) and the molecular morphology (to form percolating pathway for charge transport) do not play the significant role, opposite to the conventional viewpoint in fullerene-based cells.

In addition, our model can be straightforwardly applied to other nonfullerene cells besides the PBDB-T/ITIC blend. It is not necessary to require a NFA to have the similar A-D-A structure with that in ITIC. Proper modulation of side groups can still shape the alternating structure and large difference of electron affinity can enhance the packing of the D and A units. Rather, the parameters for efficient charge separation must be reconfirmed in other cells.

1D metal exhibits the feature of the Luttinger liquid, and one important phenomenon in 1D metals is the so-called spin-charge separation \cite{SCS,Yao08}. That is, in a non-half-filled Hubbard model with on-site repulsive interaction, it is well known that the charge excitation moves faster than spin. The present mechanism of charge separation is however different with the spin-charge separation, since we do not find the spread of holes. In addition, the electron-phonon interaction and the disorders are always essential in organic materials. Taking both the electron-phonon interactions and disorder into account, it refers to the effect of many-body localization \cite{MBL}. We expect that the many-body localization can enhance the present self-accumulation effect since in this situation the charges are more likely to be localized. This is held as the future subject.

\section*{Acknowledgment}

The authors gratefully acknowledge support from the National Natural Science Foundation of China (Grant Nos.~11574052 and 91333202). We thank Prof. Jianhui Hou and Haibo Ma for fruitful discussions.


\begin{thebibliography}{}

\bibitem{ReviewH} For a review, see J. Hou, O. Ingan\"{a}s, R. H. Friend, F. Gao, Nat. Mater. \textbf{17}, 119 (2018).

\bibitem{Chen1} L. Meng, X. Wan, L. Ding, Y. Chen et al. Science 2018. DOI:10.1126/science.aat2612.

\bibitem{NFA1} B. Kan, Q. Zhang, M. Li, X. Wan, W. Ni, G. Long, Y. Wang, X. Yang, H. Feng, and Y. Chen, J. Am. Chem. Soc. \textbf{136}, 15529 (2014).

\bibitem{NFA2} Q. Zhang, B. Kan, F. Liu, G. Long, X. Wan, X. Chen, Y. Zuo, W. Ni, H. Zhang, M. Li, Z. Hu, F. Huang, Y. Cao, Z. Liang, M. Zhang, T. P. Russell, and Y. Chen, Nat. Photon. \textbf{9}, 35 (2015).

\bibitem{Zhan2} Y. Lin, J. Wang, Z.-G. Zhang, H. Bai, Y. Li, D. Zhu and X. Zhan, Adv. Mater. \textbf{27}, 1170 (2015).

\bibitem{Zhan1} Y. Lin, Z.-G. Zhang, H. Bai, J. Wang, Y. Yao, Y. Li, D. Zhu and X. Zhan, Energy Environ. Sci. \textbf{8}, 610 (2015).


\bibitem{Zhan3} Y. Lin, Q. He, F. Zhao, L. Huo¡Í, J. Mai, X. Lu, C.-J. Su, T. Li, J. Wang, J. Zhu, Y. Sun, C. Wang, and X. Zhan, J. Am. Chem. Soc. \textbf{138}, 2973 (2016).

\bibitem{Zhan4} Y. Lin, F. Zhao, Q. He, L. Huo, Y. Wu, T. C. Parker, W. Ma, Y. Sun, C. Wang, D. Zhu, A. J. Heeger, S. R. Marder, and X. Zhan, J. Am. Chem. Soc. \textbf{138}, 4955 (2016).

\bibitem{HouAM} W. Zhao, D. Qian, S. Zhang, S. Li, O. Ingan\"{a}s, F. Gao, and J. Hou, Adv. Mater. \textbf{28}, 4734 (2016).

\bibitem{NFA3} M. Li, K. Gao, X. Wan, Q. Zhang, B. Kan, R. Xia, F. Liu, X. Yang, H. Feng, W. Ni, Y. Wang, J. Peng, H. Zhang, Z. Liang, H.-L. Yip, X. Peng, Y. Cao, and Y. Chen, Nat. Photon. \textbf{11}, 85 (2017).

\bibitem{HouJACS} W. Zhao, S. Li, H. Yao, S. Zhang, Y. Zhang, B. Yang, and J. Hou, J. Am. Chem. Soc. \textbf{139}, 7148 (2017).

\bibitem{NFA4} Y. Cui, H. Yao, Bowei Gao, Y. Qin, S. Zhang, B. Yang, C. He, B. Xu, and J. Hou, J. Am. Chem. Soc. \textbf{139}, 7302 (2017).

\bibitem{LiCM} B. Qiu, L. Xue, Y. Yang, H. Bin, Y. Zhang, C. Zhang, M. Xiao, K. Park, W. Morrison, Z.-G. Zhang, and Y. Li, Chem. Mater. \textbf{29}, 7543 (2017).

\bibitem{HouCJC} H. Yao, D. Qian, H. Zhang, Y. Qin, B. Xu, Y. Cui, R. Yu, F. Gao, and J. Hou, Chin. J. Chem. \textbf{36}, 491 (2018).

\bibitem{P14} Y. Cui, H. Yao, C. Yang, S. Zhang, and J. Hou, Acta Polymerica Sinica (2017) (doi: 10.11777/j.issn1000-3304.2018.17297).

\bibitem{Zhan5} S. Dai and X. Zhan, Acta Polymerica Sinica, \textbf{11}, 1706 (2017).



\bibitem{Ter0} X. Liu, Y. Yan, Y. Yao, and Z. Liang, Adv. Funct. Mater. \textbf{28}, 1802004 (2018).

\bibitem{Ter1} W. Su, Q. Fan, X. Guo, B. Guo, W. Li, Y. Zhang, M. Zhang, and Y. Li,  J. Mater. Chem. A, \textbf{4}, 14752 (2016).

\bibitem{Ter2} D. Baran, R. S. Ashraf, D. A. Hanifi, M. Abdelsamie, N. Gasparini, J. A. R\"{o}hr, S. Holliday, A. Wadsworth, S. Lockett, M. Neophytou, C. J. M. Emmott, J. Nelson, C. J. Brabec, A. Amassian, A. Salleo, T. Kirchartz, J. R. Durrant, and I. McCulloch, Nat. Mater. \textbf{16}, 363 (2017).

\bibitem{Ter3} W. Zhao, S. Li, S. Zhang, X. Liu, and J. Hou, Adv. Mater. \textbf{29}, 1604059 (2017).

\bibitem{Ter4} H. Zhang, X. Wang, L. Yang, S. Zhang, Y. Zhang, C. He, W. Ma, and J. Hou, Adv. Mater. \textbf{29}, 1703777 (2017).

\bibitem{Ter5} R. Yu, S. Zhang, H. Yao, B. Guo, S. Li, H. Zhang, M. Zhang, and J. Hou, Adv. Mater. \textbf{29}, 1700437 (2017).

\bibitem{Ter6} C. Wang, X. Xu, W. Zhang, S. B. Dkhil, X. Meng, X. Liu, O. Margeat, A. Yartsev, W. Ma, J. Ackermann, E. Wang, and M. Fahlman, Nano Energy \textbf{37}, 24 (2017).

\bibitem{Ter7} Y. Chen, Y. Qin, Y. Wu, C. Li, H. Yao, N. Liang, X. Wang, W. Li, W. Ma, and J. Hou, Adv. Energy Mater. \textbf{7}, 1700328 (2017).

\bibitem{Ter8} M. An, F. Xie, X. Geng, J. Zhang, J. Jiang, Z. Lei, D. He, Z. Xiao, and L. Ding, Adv. Energy Mater. \textbf{7}, 1602509 (2017).

\bibitem{Ter9} L. Zhong, L. Gao, H. Bin, Q. Hu, Z.-G. Zhang, F. Liu, T. P. Russell, Z. Zhang, and Y. Li, Adv. Energy Mater. \textbf{7}, 1602215 (2017).

\bibitem{Ter10} B. Fan, W. Zhong, X.-F. Jiang, Q. Yin, L. Ying, F. Huang, and Y. Cao, Adv. Energy Mater. \textbf{7}, 1602127 (2017).

\bibitem{Ter11} W. Zhong, J. Cui, B. Fan, L. Ying, Y. Wang, X. Wang, G. Zhang, X.-F. Jiang, F. Huang, and Y. Cao, Chem. Mater. \textbf{29}, 8177 (2017).

\bibitem{Ter12} X. A. Jeanbourquin, A. Rahmanudin, X. Yu, M. Johnson, N. Guijarro, L. Yao, and K. Sivula, ACS Appl. Mater. Interfaces \textbf{9}, 27825 (2017).

\bibitem{Ter13} N. Zhu, W. Zhang, Q. Yin, L. Liu, X. Jiang, Z. Xie, and Y. Ma, ACS Appl. Mater. Interfaces \textbf{9}, 17265 (2017).

\bibitem{Ter14} L. Yang, W. Gu, L. Hong, Y. Mi, F. Liu, M. Liu, Y. Yang, B. Sharma, X. Liu, and H. Huang, ACS Appl. Mater. Interfaces \textbf{9}, 26928 (2017).



\bibitem{Review} T. M. Clarke and J. R. Durrant, Chem. Rev. \textbf{110}, 6736 (2010).

\bibitem{Delo1} A. A. Bakulin, A. Rao, V. G. Pavelyev, P. H. M. van Loosdrecht, M. S. Pshenichnikov, D. Niedzialek, J. Cornil, D. Beljonne, and R. H. Friend, Science \textbf{335}, 1340 (2012).
\bibitem{Delo2} A. Rao, P. C. Y. Chow, S. G¨¦linas, C. W. Schlenker, C. Z. Li, H. L. Yip, A. K.-Y. Jen, D. S. Ginger, and R. H. Friend, Nature (London) \textbf{500}, 435 (2013).
\bibitem{Delo3} S. G¨¦linas, A. Rao, A. Kumar, S. L. Smith, A. W. Chin, J. Clark, T. S. van der Poll, G. C. Bazan, and R. H. Friend, Science \textbf{343}, 512 (2014).
\bibitem{Delo4} S. M. Falke, C. A. Rozzi, D. Brida, M. Maiuri, M. Amato, E. Sommer, A. De Sio, A. Rubio, G. Cerullo, E. Molinari, and C. Lienau, Science \textbf{344}, 1001 (2014).

\bibitem{Entropy}  B. A. Gregg, J. Phys. Chem. Lett. \textbf{2}, 3013 (2011).

\bibitem{Yao16} Y. Yao, X. Xie, and H. Ma, J. Phys. Chem. Lett. \textbf{7}, 4830 (2016).

\bibitem{NE} J. Liu, S. Chen, D. Qian, B. Gautam, G. Yang, J. Zhao, J Bergqvist, F. Zhang, W. Ma, H. Ade, O. Ingan\"{a}s, K. Gundogdu, F. Gao, and H. Yan, Nat. Energy \textbf{1}, 1 (2016).


\bibitem{PP1} A. De Sio, F. Troiani, M. Maiuri, J. R\'{e}hault, E. Sommer, J. Lim, S. F. Huelga, M. B. Plenio, C. A. Rozzi, G. Cerullo, E. Molinari, and C. Lienau, Nat. Commun. \textbf{7}, 13742 (2016).

\bibitem{PP2} R. Wang, Y. Yao, C. Zhang, Y. Zhang, Z. Zhang, X. Xie, H. Ma, X. Wang, Y. Li, and M. Xiao, to be published.


\bibitem {Chem} PBDB-T is the abbreviation of poly[(2,6-(4,8-bis(5-(2-ethylhexyl)thiophen-2-yl)-benzo[1,2-b:4,5-b¡¯]dithiophene))-alt-(5,5-(1¡¯,3¡¯-di-2-thienyl-5¡¯,7¡¯-bis(2-ethylhexyl)benzo[1¡¯,2¡¯-c:4¡¯,5¡¯-c¡¯]dithiophene-4,8-dione))]
and ITIC is the abbreviation of  (3,9-bis(2-methylene-(3-(1,1-dicyanomethylene)-indanone))-5,5,11,11-tetrakis(4-hexylphenyl)-dithieno[2,3-d:2¡¯,3¡¯-d¡¯]-s-indaceno[1,2-b:5,6-b¡¯]dithiophene.

\bibitem{Huang} Z. Wang, N. Zheng, W. Zhang, H. Yan, Z. Xie, Y. Ma, F. Huang, and Y. Cao, Adv. Energy Mater. \textbf{7}, 1700232 (2017).

\bibitem{Radical} Y. Li, L. Li, Y. Wu, and Y. Li, J. Phys. Chem. C \textbf{121}, 8579 (2017).

\bibitem{private} Private communications with Jianhui Hou.


\bibitem{PPnote} In polymers, the charge transfer state is more likely to be the polaron pair state \cite{PP1,PP2}. Due to the strong vibronic couplings, in small molecules we can also call the electron and hole as negative and positive charged (small) polarons. In this manner, the charge transfer state in small molecules can be regard as equivalant to the polaron pair state.

\bibitem{bridge1} H. Geng, X. Zheng, Z. Shuai, L. Zhu, and Y. Yi, Adv. Mater. \textbf{27}, 1443 (2015).

\bibitem{bridge2} G. Li, N. Govind, M. A. Ratner, C. J. Cramer, and L. Gagliardi, J. Phys. Chem. Lett. \textbf{6}, 4889 (2015).

\bibitem{Lin} S.-J. Gu, S.-S. Deng, Y.-Q. Li, and H.-Q. Lin, Phys. Rev. Lett. \textbf{93}, 086402 (2004).

\bibitem{Lut1} M. Sch\"{u}ler, M. R\"{o}sner, T. O. Wehling, A. I. Lichtenstein, and M. I. Katsnelson, Phys. Rev. Lett. \textbf{111}, 036601 (2013).

\bibitem{Lut2} H. Yoshioka, H. Shima, Y. Noda, S. Ono, and K. Ohno, Phys. Rev. B \textbf{93}, 165431 (2016).

\bibitem{SCS} C. Kollath, U. Schollw\"{o}ck, and W. Zwerger, Phys. Rev. Lett. \textbf{95}, 176401 (2005).

\bibitem{Yao08} Y. Yao, H. Zhao, J. E. Moore, and C.-Q. Wu, Phys. Rev. B \textbf{78}, 193105 (2008).

\bibitem{loss1} J. Yao, T. Kirchartz, M. S. Vezie, M. A. Faist, W. Gong, Z. He, H. Wu, J. Troughton, T. Watson, D. Bryant, and J. Nelson, Phys. Rev. Applied \textbf{4}, 014020 (2015).

\bibitem{loss2} W. Yang, Y. Luo, P. Guo, H. Sun, and Y. Yao, Phys. Rev. Applied \textbf{7}, 044017 (2017).

\bibitem{MBL} P. Bordia, H. P. L\"{u}schen, S. S. Hodgman, M. Schreiber, I. Bloch, and U. Schneider, Phys. Rev. Lett. \textbf{116}, 140401 (2016).

\end{thebibliography}
\end{document}